# First Principles Insight into Antiperovskite c-Na$_3$HS Solid State Electrolyte

Sananya Chakraborty, Nidhi Verma and Ashok Kumar*

*Department of Physics, Central University of Punjab, Bathinda, India-151401*

*Corresponding Author:  ashokphy@cup.edu.in

**Abstract**

We explore the potential of novel antiperovskite c-$Na_3HS$ to be a solid-state electrolyte for sodium-ion batteries. To investigate the dynamical stability, phase stability, thermal stability, mechanical stability and ionic, electronic and diffusive properties of c-$Na_3HS$, the first-principles methods based on density functional theory (DFT) and ab-initio molecular dynamics (AIMD) simulations have been employed. c-$Na_3HS$ has no imaginary phonon modes indicating its dynamical stability. Key findings include small energy-above-hull, the wide band gap of 4.35 eV and mechanical stability analysis that indicates the moderately hard and a little brittle nature of c-$Na_3HS$. The activation energy of Na in c-$Na_3HS$ is calculated to be ~300 meV that reduces to ~ 100 meV on introducing Na-vacancy. The ionic conductivity can be enhanced up to ~3 order of magnitude by vacancy and halogen doping in c-$Na_3HS$ structure. Thus, the obtained results indicate that c-$Na_3HS$ can be viable option to be utilized as solid-state electrolyte in sodium-ion batteries.

**Keywords:** Solid state electrolytes; Sodium-ion batteries; Antiperovskite; Density functional theory; AIMD simulations; ionic conductivity, substitutional doping

# 1. INTRODUCTION

Battery operated electric vehicles powered by renewable energy resources are considered the cleanest transportation option. The development of lithium-ion batteries (LIBs) has demonstrably revolutionized fields of communication and transportation [1]. It has been reported that LIBs have long life cycles and high energy densities [2]. However, the future of technology demands advancements beyond current LIBs. Compared to lithium, sodium is a cost-effective and earth abundant element that can be used to create large-scale rechargeable batteries in an effective and appealing way. Sodium-ion batteries (SIBs) are emerging as a substitute of LIBs with comparable performances [3].

The solid state electrolytes (SSEs) in batteries have many advantages over the liquid electrolytes such as improved safety, higher thermal and chemical stability, lower flammability and higher electrochemical stability of batteries [4, 5]. Still, some problems exist with SSEs such as electrochemical instability, dendrite growth, large-scale synthesis and interfacial resistance [6], which require further improvement. The primary characteristics of an ideal solid-state electrolyte are a large electrochemical window, high ionic conductivity, low electrical conductivity and high thermal stability [7]. Also, the material used for the solid-state electrolyte should have low activation energy and a band gap comparable to that of an insulator so that the ionic conductivity is increased and electronic conductivity is decreased [8]. Among various SSEs, inorganic halides have been considered highly stable for SSBs [9].

Also, a group of antiperovskites that are electronically inverted has been proposed [10], which was influenced by superionic conductivity of $NaMgF_3$ and $(K,Na)MgF_3$ perovskites at high temperature [10-13]. The structural modifications in the typical $ABX_3$ perovskite structure were reversed to the anti-perovskite notation $X_3BA$ according to the standard inorganic nomenclature of ionic compounds following "cation-first" approach[14], where X is now large cation (often a metal), A is Anion (often a non-metal) and B is a monovalent anion. Research interest has significantly increased concentrating on anti-perovskites for solid-state electrolytes due to properties like wide electrochemical stability window, high ionic conductivity and low cost [7, 15-17].

The recent observation of Li-rich antiperovskites has opened up a new dimension of research for solid-state electrolytes [18]. These materials showed high ionic conductivity, low melting points and three-dimensional conducting pathways. Also, $Na^+$ has larger ionic radius compared to $Li^+$ ion that helps in stabilization of the anti-perovskite structure [19]. Thus, more research is needed in studying the properties of Na-rich antiperovskites (NaRAP). Recently, a series of NaRAPs, $Na_3HCh$ (Ch= S,Se,Te) were fabricated in which both anionic sites were occupied polarizable and soft anions of $H^-$ and $Ch^{2-}$ which helps in achieving higher ionic conductivity and lower activation energy[20]. Among these $Na_3HS$ was found to exhibit an orthorhombic structure whereas the rest of all structures were cubic in nature.

Density functional theory (DFT) based First principles approach work hand-in-hand to guide the comprehension and development of battery materials such as electrolytes [21]. For example, DFT calculations on $Li_{10}GeP_2S_{12}$ have firstly clarified the 3D diffusion topology of the material [22]. These calculations can also be used to gain an insight into the performance limits, activation energy, diffusivity, ionic conductivity and potential doping techniques to modify carrier concentration [23]. Moreover, diffusion can also involve collective jumps and lattice vibrations which can be studied using ab-initio molecular dynamics (AIMD) simulations by considering all possible motions of ions and their interactions.

In this work, we investigated cubic phase c-$Na_3HS$ using DFT based first principles and AIMD simulations for solid state electrolytes. The structural, phase, dynamical, thermal and mechanical stability along with the electronic and electrochemical characteristics like - activation energy, diffusivity and ionic conductivity have been explored. Furthermore, to tune the electronic conductivity and mobility of pristine c-Na$_3$HS, different techniques like vacancy and substitutional doping have been introduced which results in enhanced diffusive an ionic property.

## 2. COMPUTATIONAL METHODOLOGY

All the calculations were executed by utilizing the Vienna Ab initio Simulation Package [24], using GGA approximation [25]. The exchange of information between the core and valence electrons was studied employing the projected augmented wave (PAW) approach[26]. An 8×8×8 Monkhorst-Pack k-point grid, a plane wave cutoff energy of 450 eV and 10$^{-5}$ eV energy convergence are used in the calculations. The conjugate gradient approach was implemented to minimize all energy until the Hellman–Feynman forces performing on all atoms were less than 10$^{-3}$ eV/Å. For sodium, hydrogen and sulphur atoms, the pseudopotentials having valence configurations of 2p$^6$3s$^1$, 1s$^1$ and 3s$^2$3p$^4$, respectively was used. To determine the dynamic stability of c-Na$_3$HS, a phonon dispersion plot with a q-point mesh of 2×2×2 was obtained using density functional perturbation theory [27]. To ensure an accurate calculation of electronic band structure of c-Na$_3$HS, hybrid functional Heyd-Scuseria-Ernzerhof (HSE06) was used[28, 29], as the conventional DFT is unable to estimate the band gap accurately. The thermal stability has been analysed via AIMD simulations within the NVT ensemble at 1000 K.

To investigate the diffusivity, activation energy and ionic conductivity, AIMD simulations were performed on a 3×3×3 supercell with a total energy convergence of 10$^{-4}$ eV. The total simulation time of the AIMD simulations was 50ps; with 2fs time steps. All simulations were implemented in the NVT ensemble while employing the Nose-Hover thermostat at 4 temperatures: 400K, 600K, 800K and 1000K. To analyse diffusive and ionic properties, Python Materials Genomics (PyMatGen) library [30] has been used. VASPKIT [31] is used for generating K-points for the band structure and mechanical stability calculations.

## 3. RESULTS AND DISCUSSIONS
### 3.1 Structure and stability analysis

c-$Na_3HS$ is an antiperovskite structure as depicted in Figure 1(a) that crystallizes in the cubic Pm-3m space group. The relaxed c-$Na_3HS$ unit cell holds five atoms having a lattice constant 4.39 Å. S atom is bonded to 12 equivalent Na atoms to form $SNa_{12}$ cuboctahedra that share corners with 12 equivalent $SNa_{12}$ cuboctahedra and, faces with eight equivalent $HNa_6$ octahedra. Four comparable S atoms and two analogous H atoms are bound to Na atom in a linear geometry. H atom is linked to six equivalent Na atoms to form $HNa_6$ octahedra and faces 8 equivalent $SNa_{12}$ cuboctahedra.

The structural stability of material can also be assessed theoretically using the phase diagram which can be built based on free energy. The phase stability of material with a clearly defined chemical composition is evaluated using the energy above the hull, or $E_{hull}$. A stable compound often possesses an $E_{hull}$ of 0 or lower while a higher $E_{hull}$ indicates that it is likely to be unstable [32]. In order to create a phase diagram, a suitable free energy model for the system of choice must be used to compare the relative thermodynamic stability of the phases that make up the system [33]. For an isothermal closed Na-H-S system, Gibbs free energy, G is the most appropriate thermodynamic potential, which is a Legendre transform of the internal energy, E and enthalpy, in the following way: [34]

$$G(T, P, N_{Na}, N_H, N_S) = H(T, P, N_{Na}, N_H, N_S) - TS(T, P, N_{Na}, N_H, N_S) = E(T, P, N_{Na}, N_H, N_S) + PV(T, P, N_{Na}, N_H, N_S) - TS(T, P, N_{Na}, N_H, N_S) \quad (1)$$

Where, T represents temperature, P is pressure, S depicts entropy, V denotes volume and $N_i$ is the number of atoms in the system of species Na. Here, PV term can be neglected as PΔV is small. Therefore, the expression of G reduces to E at 0K temperature.

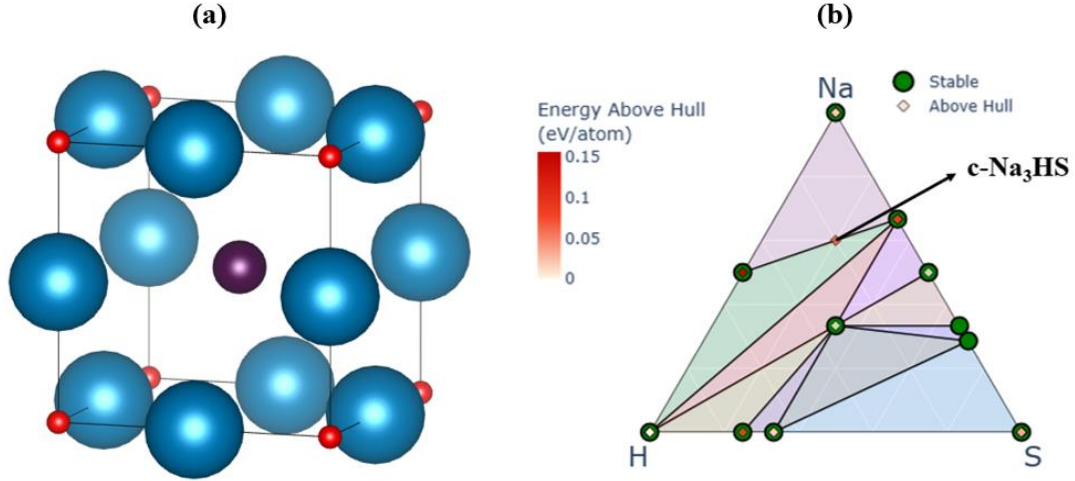

**Figure 1:** **(a)** c-Na$_3$HS structure. Na$^+$ is in blue, H$^-$ is in red and S$^{-2}$ is in purple and **(b)** Ternary phase diagram of Na$_3$HS. The stable phases with green dots and energy above Hull for c-Na$_3$HS are also shown.

For c-Na$_3$HS, the energies of most of the reported phases were calculated in the Na-H-S chemical space. Next, the convex hull of four-dimension (E, xNa, xH, xS) space was formed at 0 K to create the phase diagram, where x$_i$ is the atomic portion of element X and E is the normalised energy per atom. Now, in the compositional coordinate space of Na-H-S, the phase diagram is produced by projecting the vertices on the convex hull. From the ternary phase diagram of c-Na$_3$HS (Figure 1(b)), the energy above hull was calculated to be 78 meV/atom, which indicates the cubic phase of Na$_3$HS is energetically stable.

**Thermal and dynamical stability**

We now investigated the thermal stability of c-Na$_3$HS at 1000 K for 15 ps (Figure 2(a)). The energy versus time steps plot shows that the energy undergoes marginal fluctuations in its value and the geometry of the c-Na$_3$HS structure is slightly distorted without breaking of bonds which suggests that c-Na$_3$HS maintains its thermal stability at high temperature.

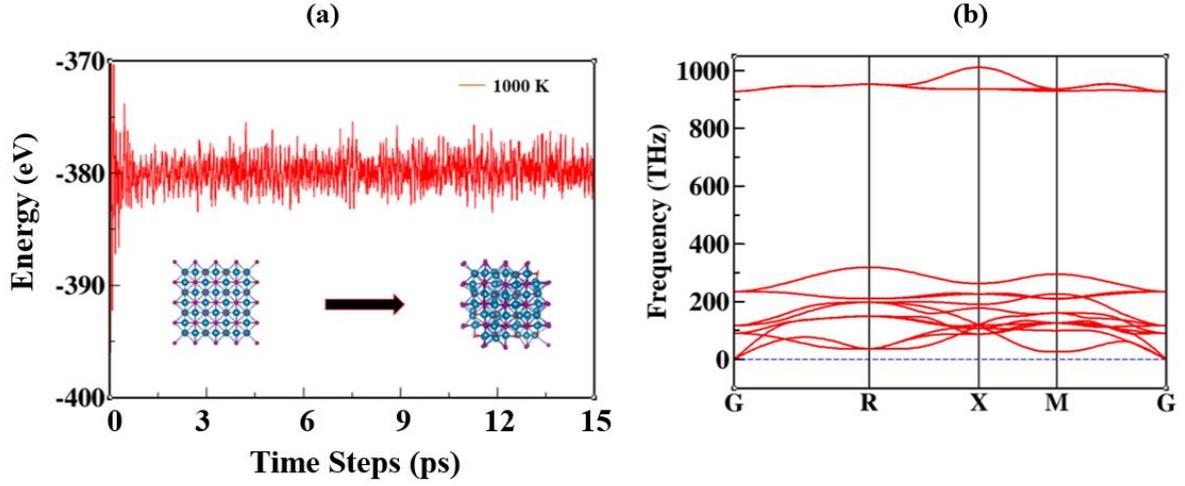

**Figure 2:** **(a)** Energy fluctuations during AIMD simulations with initial and final structure at 1000 K after 15ps **(b)** Phonon dispersion curves of c-Na$_3$HS.

Next, we analysed the dynamical stability by considering the phonon dispersion plot in the Brillouin zone with respect to high-symmetry points. The calculated phonon spectrum shows that the imaginary frequencies are absent throughout the whole Brillouin zone confirming the dynamic stability of structure c-Na$_3$HS.

**Mechanical Stability.** In the design of a solid state electrolyte, ensuring mechanical stability and evaluating the mechanical properties are essential, as it affect the battery manufacturing, operation and performance[9]. The mechanical stability is investigated by determining whether the Born stability criteria for cubic systems [35] is satisfied or not. It is observed that the necessary and sufficient Born stability criteria is satisfied:

$$C_{11}-C_{12} > 0 \; ; \; C_{11}+2C_{12} > 0 \; ; \; C_{44} \qquad (2)$$

This establishes the fact that c-Na$_3$HS exhibits mechanically stability. Various mechanical properties such as Young's modulus (Y), Bulk modulus (B), Shear modulus (G) and Pugh's ratio (B/G) have also been calculated. A comparative analysis with some of the commonly used sodium-ion solid-state electrolyte materials is given in Table 1.

**Table 1.** Young's Modulus (Y), Bulk Modulus (B), Shear Moduli (G), and Pugh's Ratio (B/G) of c-Na$_3$HS compared with common sodium-ion solid state electrolyte materials.

| Materials | Y (GPa) | B (GPa) | G (GPa) | B/G |
|---|---|---|---|---|
| c-Na$_3$HS (this work) | 57.18 | 30.83 | 24.01 | 1.29 |
| NASICON (NaZr$_2$(PO$_4$)$_3$) [36] | 120.9 | 86.3 | 47.7 | 1.82 |
| Na$_3$OBr [36] | 57.4 | 34 | 23.6 | 1.45 |
| Na$_3$OCl [36] | 60.2 | 36.4 | 24.6 | 1.47 |
| Na$_3$PS$_4$ (cubic) [36] | 32.6 | 21.5 | 13.1 | 1.64 |

It is evident from Table 1 that c-Na$_3$HS exhibits lower values of B, G and Y as compared to NASICON and Na$_3$OCl but higher than Na$_3$PS$_4$. Also, these values are comparable to that of Na$_3$OBr. This observation places c-Na$_3$HS in the category of materials having moderate hardness. This level of hardness is sufficient to withstand any external stresses in solid-state SIBs. On the other hand, a certain degree of softness is required in the material to achieve maximum electrode contact [9]. c-Na$_3$HS has lower values of Y than NASICON, Na$_3$OCl and Na$_3$OBr indicating that it has a certain degree of softness. The Pugh's ratio underscores the brittle or ductile nature of a material. Any material having Pugh's ratio less than 1.75 can show brittle nature [37]. From Table 1 it is observed that NASICON is truly ductile in nature. c-Na$_3$HS is observed to have the lowest Pugh's ratio among other reported structures, which makes it slightly more brittle than other materials mentioned. The hardness and certain degree of softness in c-Na$_3$HS makes it suitable for consideration for solid-state electrolyte material.

### 3.2 Electronic Structure

The electronic band structure of c-Na$_3$HS is calculated in the Brillouin zone, using more accurate description of exchange-correlation in the calculations [38], along high-symmetry points. The electronic band structure of c-Na$_3$HS (Figure 3) reveals that it has an indirect band gap of 4.25 eV (HSE06 functional) and 3.2 eV when calculated using GGA-PBE functional with conduction band minima at X and valence band maxima at M point. Note that the HSE06 hybrid functional shows a more accurate band gap than the GGA functional for the materials having a band gap between 1-5 eV [39]. The projected density of states (PDOS) and total density of states (TDOS) are provided in Figure S1, ESI. The states near Fermi level in valance band and conduction band side are primarily contributed by S and Na atoms, respectively. The band gap of c-Na$_3$HS establishes it to be suitable candidate for solid-state electrolyte in SIBs.

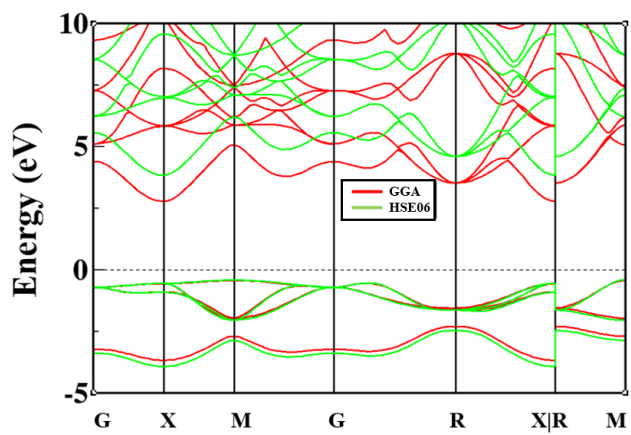

**Figure 3:** Electronic band structure of c-Na$_3$HS employing HSE06 and GGA functional.

**3.3 Diffusive and Ionic properties**.

For designing an efficient solid-state electrolyte, diffusivity, activation energy and ionic conductivity are crucial parameters which have been investigated by performing AIMD simulations on the supercell of the c-Na$_3$HS structure. We assessed the diffusion properties of Na-ions, by computing the self-diffusion coefficient of sodium ions based on AIMD trajectories. Mean square displacement (MSD) for Na-ions in the electrolyte c-Na$_3$HS materials is calculated at 400K, 600K, 800K and 1000 K, within a 50 ps time step window (Figure 4 (a)). It is evident that the MSD of Na increases slowly at lower temperatures but at 1000 K it starts showing a linear relationship with time steps. The MSD plots hints that the diffusivity is very low at lower temperatures but increases at 1000 K. Note that the diffusivity of Na in c-Na$_3$HS can be determined from the relation:

$$D = \frac{1}{2dt} < [\Delta r(t)]^2 > \qquad (3)$$

Where D is diffusion coefficient or diffusivity, dimensionality factor is equivalent to 3 for c-Na$_3$HS as cubic Na$_3$HS is a bulk structure and $< [\Delta r(t)]^2 >$ represents average value of MSD concerning time duration of t. From Figure 4(a), it can be comprehended that there is very less diffusion of Na-atoms as the atoms move around their equilibrium position mostly. Also, the diffusion along different axis of c-Na$_3$HS remain nearly the same (Figure S2, ESI).

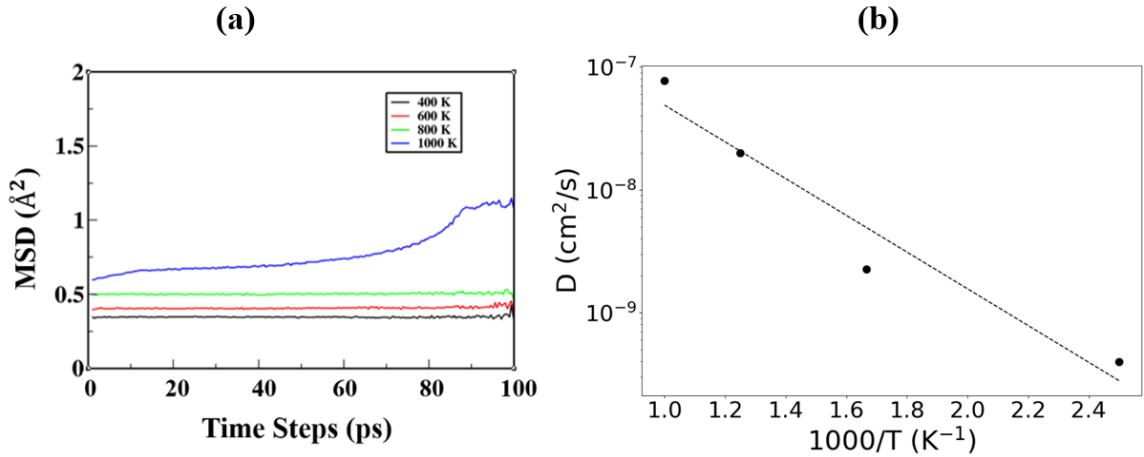

**Figure 4.** (a) Mean Square displacement of Na for c-Na$_3$HS at 400K, 600K, 800K and 1000 K and (b) Arrhenius plot of c-Na$_3$HS.

Assuming the absence of phase transitions and plenty of mobile defect carriers, an Arrhenius relationship [40] can be used for the diffusivity in a solid-state electrolyte material as:

$$D = D_0 \exp\left(\frac{-E_a}{K_B T}\right) \qquad (4)$$

Where maximum diffusivity at infinite temperature is denoted by $D_0$, $K_B$ is the Boltzmann constant, $E_a$ depicts activation energy, T is the temperature. Thus, $E_a$ can be calculated by modeling the Arrhenius curve obtained by plotting the values of diffusion coefficients (cm$^2$/sec) versus 1000/T (K$^{-1}$) (Figure 4 (b)). The slope obtained from fitting of the plot gives the activation energy. The activation energy of Na$^+$ in c-Na$_3$HS is calculated to be 0.298 $\pm$ 0.05 eV.

Furthermore, to determine the extrapolated ionic conductivity of Na$^+$ at room temperature, Nernst-Einstein equation [41] is used:

$$\sigma = \frac{\rho F^2 z^2}{RT} D \quad (5)$$

Where, F represents Faraday constant, $\rho$ is molar density, z shows valence charge of Na-ion ($z = +1$), and gas constant is denoted be R. The diffusion coefficient is extrapolated at 300 K and then the extrapolated ionic conductivity is calculated. The ionic conductivity is 0.01 mS/cm at 300 K which is low than previously reported solid-state electrolyte materials (Table 2). It can be also inferred from Table 2 that the activation energy of c-Na$_3$HS is lower than Na$_2$(BH$_4$)(NH$_2$), Na$_7$P$_3$O$_{11}$, NASICON, orthorhombic Na$_3$HS and Na$_3$La$_5$Cl$_{18}$.

**Table 2.** The assessment of solid-state electrolyte materials concerning activation energy and ionic conductivities different for sodium-ion batteries at 300 K.

| Material | Activation energy (meV) | Ionic Conductivity (mS/cm) |
|---|---|---|
| c-Na$_3$HS (this work) | 298 | 0.01 |
| Na$_2$(BH$_4$)(NH$_2$) [42] | 590 | 0.003 |
| Na$_7$P$_3$O$_{11}$ [43] | 535 | 0.003 |
| Na$_6$SbS$_5$Cl [44] | - | 0.022 |
| NASICON [45] | 302 | 0.1 |
| Na$_3$PS$_4$ (cubic) [46] | 200 | 0.46 |
| Na$_3$SbS$_4$ [47] | 220 | 1 |
| Na$_3$HS (orthorhombic) [20] | 300 | - |
| Na$_3$SO$_4$H [48] | - | 4 x 10$^{-4}$ |
| Na$_3$La$_5$Cl$_{18}$ [49] | 320 | 1.30 |
| Na$_3$OBH$_4$ [50] | 250 | 4.4 |

**3.4 Ionic Conductivity enhancement via single ion vacancy**

For a notable enhancement in the ionic conductivity of c-Na$_3$HS, vacancy engineering is one of the most researched and successful approaches. This approach tailors a variety of material properties, such as distribution of charges, electronic properties, and surface stability. We now explore the effect of introducing vacancy in c-Na$_3$HS on the ionic conductivity and activation energy. The vacancy is introduced by removing single Na-atom from a 3×3×3 supercell of c-Na$_3$HS which leads to the composition of c-Na$_{2.96}$HS. This composition is optimized with no change in the lattice parameters of the cell. MSDs were calculated for Na$_{2.96}$HS composition at 400, 600, 800 and 1000 K, within a 50 ps window. From Figure 5, it is evident that the MSDs increase linearly with time and also with increasing temperatures. The MSD plot increases linearly with time for c-Na$_{2.96}$HS as compared to nearly flat MSD curve for pristine c-Na$_3$HS (Figure S3, ESI). For the composition Na$_{2.96}$HS, the MSD plots hint at the increased movement of Na-ions leading to higher values of diffusion coefficients and hence higher ionic conductivity. From the Arrhenius plot (Figure 5(b)), the activation energy, $E_a$, for Na$_{2.96}$HS is evaluated to be 0.12 ± 0.0067 eV, which is lower than the pristine c-Na$_3$HS (0.298 ± 0.05 eV).

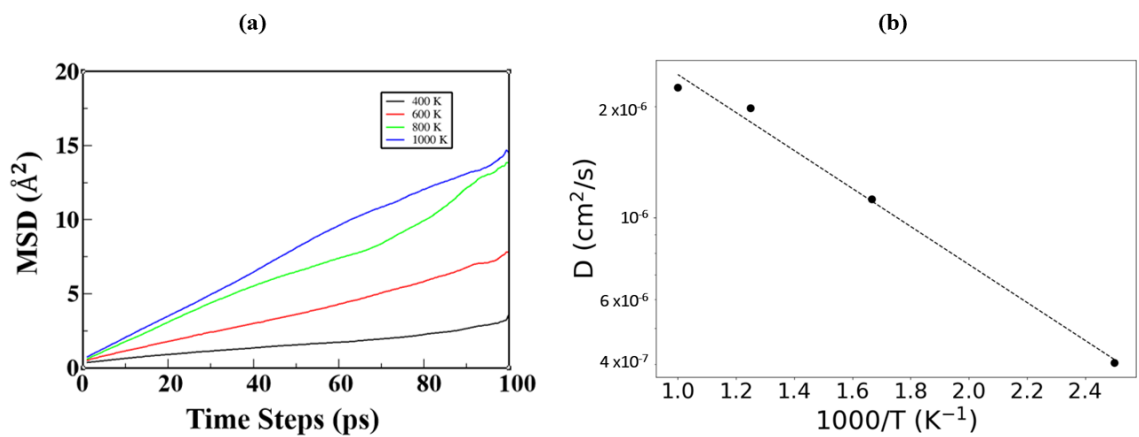

**Figure 5**: **(a)** Mean Square displacement of Na for c-Na$_{2.96}$HS composition after introducing single Na-ion vacancy at 400, 600, 800 and 1000 K and **(b)** the corresponding Arrhenius plot.

We have also compared the calculated ionic conductivity of pristine c-$Na_3HS$ and $Na_{2.96}HS$ at different temperatures which have been given in Table 3. The ionic conductivity is calculated using the extrapolated diffusivity at 300 K, which comes out to be 33.19 mS/cm for single vacancy-defected $Na_{2.96}HS$ as compared to 0.01 mS/cm for pristine c-$Na_3HS$. Thus, the obtained ionic conductivity is of the order of ~ 3 magnitude larger than the pristine c-$Na_3HS$'s ionic conductivity of (Table 3). Thus, it can be inferred that introducing vacancy enhances the ionic conductivity of pristine c-$Na_3HS$ that results into increase in performance of solid-state electrolyte for SIBs.

**Table 3:** Comparison of ionic conductivities at different temperatures for pristine c-$Na_3HS$ and single vacancy-defected $Na_{2.96}HS$.

| Temperature (K) | Ionic Conductivity (mS/cm) | |
|---|---|---|
| | c-$Na_3HS$ | c-$Na_{2.96}HS$ |
| 300 | 0.01 | 33.19 |
| 400 | 0.04 | 66.93 |
| 600 | 0.31 | 119.94 |
| 800 | 0.76 | 147.49 |
| 1000 | 1.24 | 158.74 |

**3.5 Enhancement of Ionic conductivity via substitutional halogen doping.**

Along with vacancies, we also explored the doping strategy which alters charge carrier concentration of pristine structure leading to the formation of ionisable entities inside structure that affect mobility and conductivity. The vacancies can also be introduced via substitutional doping strategy which requires charge compensation. In this case, one Cl atom took the place of one S atom and for charge compensation one Na atom was removed from a 3×3×3 supercell of c-$Na_3HS$. This led to the composition of $Na_{2.96}HS_{0.96}Cl_{0.04}$. The AIMD simulations on single Cl atom doped structure $Na_{2.96}HS_{0.96}Cl_{0.04}$ at 600 K and 1000 K temperatures show a possibility of enhancement of ionic conductivity (Figure S4, ESI).

Introducing more than one vacancy is expected to increase the ionic conductivity further. Therefore, in a 3×3×3 supercell we substitute two S atoms with two X (= Cl, Br) atoms and two Na atoms are removed keeping in mind the charge compensation. The resulting configuration was $Na_{2.92}HS_{0.92}X_{0.07}$ (X = Cl, Br). The AIMD simulations at 600 K were executed within 50 ps for this composition and the ionic conductivity was compared to pristine c-$Na_3HS$. The MSD plots in Figure 6 and Figure S5, ESI provide a hint towards the increased diffusivity followed by an increase in ionic conductivity. On calculating the ionic conductivity at 600 K, it was observed that, the ionic conductivity for Br atom doped composition ($Na_{2.92}HS_{0.92}Br_{0.07}$) is 148.35 mS/cm and that for the Cl atom doped composition ($Na_{2.92}HS_{0.92}Cl_{0.07}$) is 137.97 mS/cm which is quite higher than pristine c-$Na_3HS$ (0.31 mS/cm). It is evident that ionic conductivity of pristine c-$Na_3HS$ is less than some sodium-ion-based solid-state electrolytes (Table 2) but after incorporating the vacancies and halogen doping, there is an approximately three-order magnitude improvement in ionic conductivity.

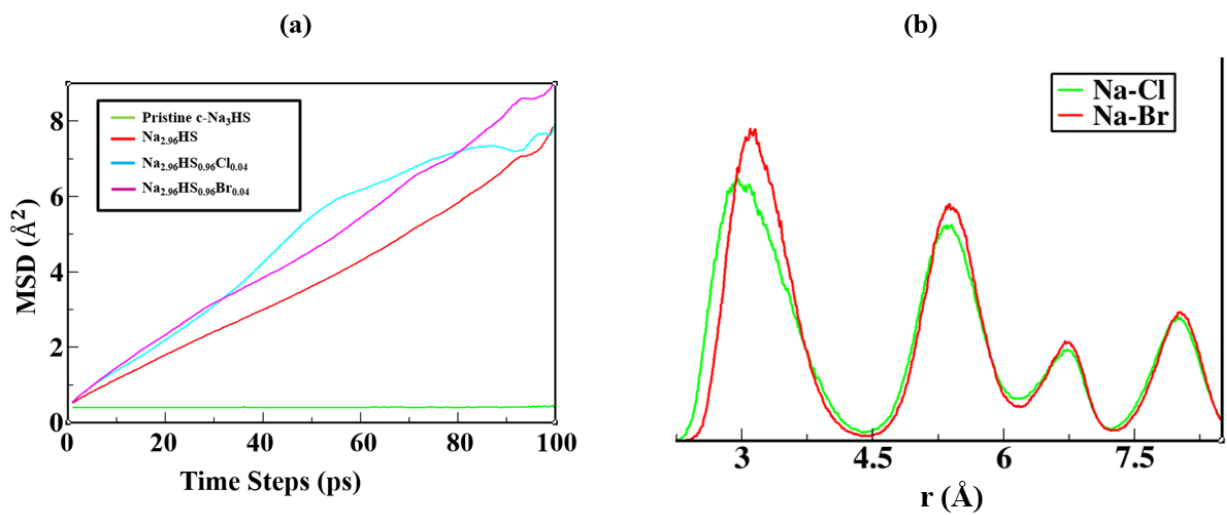

**Figure 6**: (a) MSD plots of pristine c-$Na_3HS$, $Na_{2.96}HS$, $Na_{2.92}HS_{0.92}Cl_{0.07}$ and $Na_{2.92}HS_{0.92}Br_{0.07}$ at 600 K. (b) Radial distribution functions (RDFs) for Na-Cl distribution in $Na_{2.92}HS_{0.92}Cl_{0.07}$ and Na-Br distribution in $Na_{2.92}HS_{0.92}Br_{0.07}$.

From Figure 6(a), it is evident that the ionic conductivity is enhanced as the ionic radius of the halogen atom is increased from Cl to Br. To elucidate this more, the radial distribution functions (RDFs) for Cl and Br atoms are shown in Figure 6 (b). On increasing the ionic radius from Cl to Br, the first peak of Na-distributions becomes narrower indicating the Na positions are well-defined in the structure doped with Br than in the Cl-doped structure. The shift in the first peak is due to the increased ionic radius of Br atom. To achieve future-focused solid electrolytes, it is essential to comprehend the mechanisms that both enhance and impede sodium-ion conductivity

## 4. CONCLUSIONS

In summary, first principles DFT and AIMD simulations have been utilized to explore the ability and suitability of c-$Na_3HS$ to be regarded as potential solid-state electrolyte for sodium-ion batteries. The stability analysis shows good thermal and dynamical stability of c-$Na_3HS$ with minimum energy fluctuations and lack of any imaginary frequencies in plot of phonon dispersion, respectively. The phase stability analysis established that c-$Na_3HS$ is energetically stable with energy-above hull of 78 meV/atom. Moreover, the mechanical stability hints at the material's moderate hardness, which is desired in a solid-state electrolyte material. The AIMD simulations performed on pristine c-$Na_3HS$ help to obtain understanding about the diffusive and ionic properties of c-$Na_3HS$. The study concludes that the c-$Na_3HS$ can be a potential solid-state electrolyte by enhancing its properties via vacancies and doping halogen atoms in the structure of sodium-ion batteries.

**ACKNOWLEDGEMENTS**

The results presented in this paper are obtained using the computational facility available at the Central University of Punjab, Bathinda. NV is grateful to CSIR for awarding the Senior Research Fellowship (SRF).